\documentclass[twocolumn,showpacs,superscriptaddress]{revtex4-1}
\usepackage{mymacros}
\usepackage{graphicx}
\usepackage{units}
\usepackage{multirow}
\usepackage{array}
\usepackage{amsfonts}
\usepackage{amsmath}
\usepackage{amssymb}
\usepackage{amsfonts}
\usepackage{epstopdf}
\usepackage{textcomp}
\usepackage{color}
\usepackage{subfigure}
\usepackage{xspace}
\RequirePackage{lineno}
\begin{document}

\title{\boldmath Measurement of absolute branching fraction of the inclusive decay $\lcptolx$ }

\author{
\begin{small}
\begin{center}
M.~Ablikim$^{1}$, M.~N.~Achasov$^{9,d}$, S.~Ahmed$^{14}$, M.~Albrecht$^{4}$, M.~Alekseev$^{53A,53C}$, A.~Amoroso$^{53A,53C}$, F.~F.~An$^{1}$, Q.~An$^{50,40}$, J.~Z.~Bai$^{1}$, Y.~Bai$^{39}$
% ...
\\
\vspace{0.2cm}
(BESIII Collaboration)\\
\vspace{0.2cm} {\it
$^{1}$ Institute of High Energy Physics, Beijing 100049, People's Republic of China\\
% ...
$^{4}$ Bochum Ruhr-University, D-44780 Bochum, Germany\\
% ...
$^{9}$ G.I. Budker Institute of Nuclear Physics SB RAS (BINP), Novosibirsk 630090, Russia\\
% ...
$^{14}$ Helmholtz Institute Mainz, Johann-Joachim-Becher-Weg 45, D-55099 Mainz, Germany\\
% ...
$^{39}$ Southeast University, Nanjing 211100, People's Republic of China\\
$^{40}$ State Key Laboratory of Particle Detection and Electronics, Beijing 100049, Hefei 230026, People's Republic of China\\
% ...
$^{50}$ University of Science and Technology of China, Hefei 230026, People's Republic of China\\
% ...
$^{53}$ (A)University of Turin, I-10125, Turin, Italy; (B)University of Eastern Piedmont, I-15121, Alessandria, Italy; (C)INFN, I-10125, Turin, Italy\\
% ...
$^{d}$ Also at the Novosibirsk State University, Novosibirsk, 630090, Russia\\
% ...
}\end{center}
    \vspace{0.4cm}
\end{small}
}
\noaffiliation{}
%%% Local Variables:
%%% mode: latex
%%% TeX-master: "omega-chicj"
%%% End:

%\vspace{...}
%%%%%%%%%%%%%%%%%%%%%%%%%%%%%%%%%%%%%%%%%%%%%%%%%%%%%%%%%%%%%%%%
%%%%%     Abstract           Part                  %%%%%%%%%%%%%
%%%%%%%%%%%%%%%%%%%%%%%%%%%%%%%%%%%%%%%%%%%%%%%%%%%%%%%%%%%%%%%%

\begin{abstract}
\vspace{0.5cm}
Based on an $e^+e^-$ collision data sample corresponding to an
integrated luminosity of 567$\,\rm{pb}^{-1}$ taken at the center-of-mass energy of $\sqrt{s} = 4.6$\,GeV with the BESIII detector, we measure the absolute  branching fraction of the inclusive decay $\Lambda_{c}^{+} \to \Lambda + X$ to be $\mathcal{B}(\Lambda_{c}^{+} \to \Lambda + X)=(38.2^{+2.8}_{-2.2}\pm0.8)\%$ using the double-tag method, where $X$ refers to any possible final state particles. In addition, we search for direct \CP violation in the charge asymmetry of this inclusive decay for the first time, and obtain $\mathcal{A}_{\CP} \equiv \frac{\mathcal{B}(\Lambda_{c}^{+} \to \Lambda + X)-\mathcal{B}(\bar{\Lambda}_{c}^{-} \to \bar{\Lambda} + X)}{\mathcal{B}(\Lambda_{c}^{+} \to \Lambda + X)+\mathcal{B}(\bar{\Lambda}_{c}^{-} \to \bar{\Lambda} + X)} = (2.1^{+7.0}_{-6.6}\pm1.4)\%$, a statistically limited result with no evidence of \ensuremath{C\!P}\xspace violation.
\end{abstract}

\maketitle

%\linenumbers

%%%%%%%%%%%%%%%%%%%%%%%%%%%%%%%%%%%%%%%%%%%%%%%%%%%%%%%%%%%%%%%%
%%%%%     Introduction       Part                  %%%%%%%%%%%%%
%%%%%%%%%%%%%%%%%%%%%%%%%%%%%%%%%%%%%%%%%%%%%%%%%%%%%%%%%%%%%%%%

The inclusive decay $\lcptolx$, where $X$ means any possible final state particles, is mediated by the $c \to s$ Cabibbo-favored (CF) transition that dominates the decays of the $\lcp$~\cite{Korner:1978ec,Asner:2008nq,Cheng:2015iom}.
As the $\Lambda^+_c$ is the lightest charmed baryon, the decay rate of the $\lcptolx$ is important to calibrate the amplitude of the CF transition in the charmed baryon sector in theory, which suffers from a large uncertainty in the non-perturbative QCD region~\cite{Cheng:2015iom}.
For instance, the $\lcptolx$ decay rate is an essential input in the calculation of the lifetimes of charmed baryons, whose current theoretical results largely deviate from the experimental measurements~\cite{Cheng:1997xba, Cheng:2015iom, pdg1}.
Furthermore, better understanding of the quark structure and decay dynamics in the $\lcptolx$ benefits the research on heavier charmed baryons~\cite{Cheng:1991sn,Cheng:1993gf}. Especially for those lesser-known charmed baryons with double- or triple-charm quarks, an improved and calibrated theoretical prediction on the $c\to s$ decay vertex is crucial for guiding experimental search~\cite{Yu:2017zst, Geng:2017mxn}, such as the observation of the $\Xi_{cc}^{++}$ at LHCb~\cite{,Aaij:2017ueg}.

Measurements of the branching fraction (BF) of this decay were carried
out only before 1992 by the SLAC Hybrid Facility Photon, Photon
Emulsion and CLEO collaborations~\cite{exp1,exp2,exp3}. The average of
their results gives  ${\mathcal{B}} (\lcptolx) =
(35\pm11)\%$~\cite{pdg1}, with an uncertainty larger than 30\%.  The
three individual  measurements show big discrepancies, and their
average in the Particle Data Group (PDG) gives a poor fit quality of
$\chi^2/\text{ndf}=4.1/2$ and a low confidence level of
0.126~\cite{pdg1}. This is because they were not absolute measurements
and substantial uncertainties could be underestimated. Hence, it is
crucial to carry out an absolute measurement with improved
precision. Furthermore, the sum of the BFs of the known exclusive
decay final states involving the $\Lambda$ in PDG is $(24.5 \pm
2.1)\%$~\cite{pdg1}. The difference between the inclusive and
exclusive rates will point out the size of as yet unknown decays,
which requires high precision measurement of ${\mathcal{B}}
(\lcptolx)$. In addition, precise knowledge of ${\mathcal{B}}
(\lcptolx)$ provides an essential input for exploring the decays of
$b$-flavored hadrons involving a $\lcp$ in the final states.

It has been confirmed that the Cabibbo-Kobayashi-Maskawa (CKM)
mechanism embedded in the Standard Model (SM) is the main source of
\CP violation in the quark sector~\cite{ckm2}. The impressive
agreement on \CP violation among the results from the $s$-quark and
$b$-quark sectors~\cite{Charles:2004jd,Bona:2005vz}, calls for further
checks in the less tested area of $c$-quark sector. The SM predictions
for \CP violation in the charm sector are tiny due to the hierarchical
structure of the CKM matrix and the mass differences between the
fermion generations. Any significant amount of \CP violation would be
an observation of physics beyond the SM, and therefore, the charmed baryon decays provide an opportunity to improve our knowledge on \CP violation in and beyond the SM~\cite{cpcharm1,cpcharm2,cpcharm3,cpcharm4}. In this analysis, we search for direct \CP violation by measuring the charge asymmetry of this inclusive decay
$\mathcal{A}_{\CP} \equiv \frac{{\mathcal{B}}(\lcptolx)-{\mathcal{B}}(\lcmtolx)}{{\mathcal{B}}(\lcptolx)+{\mathcal{B}}(\lcmtolx)}$.

The data used in this Letter comprise an integrated luminosity of
$567\,\rm{pb}^{-1}$~\cite{llpair}, corresponding to about $1.0 \times
10^{5}$  $\lcp\lcm$ pairs~\cite{Ablikim:2015flg}. The data set was
collected with the BESIII detector  at the center-of-mass energy
$\sqrt{s} = 4.6$~GeV. At this energy, the $\lcp\lcm$ pairs are
produced near the production threshold with no additional hadrons,
providing a clean environment for studying $\lcp$ decays. By analyzing
the data with the double-tag (DT) method~\cite{mark3}, we perform the
first measurement of the absolute BF for the inclusive decay
$\lcptolx$. Throughout this Letter, charge-conjugate modes are
implicitly assumed, unless explicitly stated.

Details about the features and capabilities of the BESIII detector can
be found in Ref.~\cite{Ablikim:2009aa}.  The response of the
experimental apparatus is simulated with a
{\sc{geant4}}-based~\cite{Agostinelli:2002hh} Monte Carlo\,(MC)
simulation package. The reactions in $e^+e^-$ annihilations are
generated by {\sc{kkmc}}~\cite{Jadach:1999vf} and
{\sc{evtgen}}~\cite{Lange:2001uf}, with initial-state radiation\,(ISR)
effects~\cite{Kuraev:1985hb} and final-state radiation\,(FSR)
effects~\cite{RichterWas:1992qb} included. To study backgrounds,
optimize event selection criteria and validate data analysis method,
an inclusive MC sample is produced at $\sqrt{s} = 4.6 \gev$. This
sample consists of pair production of charmed mesons ($D$ and $D_{s}$)
and baryons ($\lcp$), the ISR-produced $\psi$ states and quantum
electrodynamics processes.  The $\lcp$ is set to decay to all
possible final states based on the BFs from the Particle Data Group
(PDG)~\cite{pdg2}.

%%%%%%%%%%%%%%%%%%%% // Analysis // %%%%%%%%%%%%%%%%%%%%%%%
To study the signal $\lcptolx$ with the DT method, we first select the anti-particle $\lcm$ by the two decay modes, $\lcmtopks$ and $\lcmtopkpi$. The yield of the tag mode $i$, $N^{\rm{tag}}_{i}$, is given by
\begin{linenomath*}
\begin{equation}
N^{\rm{tag}}_{i} = 2 \cdot N_{\lcp\lcm} \cdot \mathcal{B}^{\rm{tag}}_{i} \cdot \varepsilon^{\rm{tag}}_{i},
\label{equ:ntag}
\end{equation}
\end{linenomath*}
where $N_{\lcp\lcm}$ is the number of $\lcp\lcm$ pairs in the data sample, while $\mathcal{B}^{\rm{tag}}_{i}$ and $\varepsilon^{\rm{tag}}_{i}$ are the BF and detection efficiency for the tag mode $i$. Then we search for a $\lmd$ among the remaining tracks. The number of the inclusive decays of $\lcptolx$ in the presence of the tag mode $i$, $N^{\rm{sig}}_{i}$, is given by
\begin{linenomath*}
\begin{equation}
N^{\rm{sig}}_{i} = 2 \cdot N_{\lcp\lcm} \cdot \mathcal{B}^{\rm{tag}}_{i} \cdot \mathcal{B}^{\rm{sig}} \cdot \varepsilon^{\rm{tag}}_{i} \cdot \varepsilon^{\rm{sig}},
\label{equ:nsig}
\end{equation}
\end{linenomath*}
where  $\mathcal{B}^{\rm{sig}}$ and $\varepsilon^{\rm{sig}}$ are the BF and reconstruction efficiency for the inclusive decay $\lcptolx$. Here we assume that the reconstruction efficiency $\varepsilon^{\rm{sig}}$ is independent of the tag mode, so the DT efficiency is given by $\varepsilon^{\rm{sig,tag}}_{i} \approx \varepsilon^{\rm{sig}} \cdot \varepsilon^{\rm{tag}}_{i}$. From Eq.~(\ref{equ:ntag}) and Eq.~(\ref{equ:nsig}) we determine the BF of the signal process by
\begin{linenomath*}
\begin{equation}
\mathcal{B}^{\rm{sig}}=\frac{(\sum_{i}N^{\rm{sig}}_{i})/\varepsilon^{\rm{sig}}}{\sum_{i}N^{\rm{tag}}_{i}}.
\label{equ:bsigs}
\end{equation}
\end{linenomath*}

The efficiency for detecting $\lmd$ as a function of momentum and
polar angle is determined from data with a control sample of $J/\psi$ decays.
Due to lacking knowledge of the phase space distribution of the inclusive decay $\lcptolx$, we re-weight the $\lmd$ efficiencies
according to the momentum and polar angle distributions of $\lmd$ in the DT signals.
Therefore, the signal BF is calculated by
\begin{linenomath*}
\begin{equation}
\mathcal{B}^{\rm{sig}}=\frac{\sum_{j} ((\sum_{i}{N^{\rm{sig}}_{i,j})/\varepsilon^{\rm{sig}}_{j}})}{\sum_{i}N^{\rm{tag}}_{i}}
=\frac{\sum_{j}(N^{\rm{sig}}_{-,j}/\varepsilon^{\rm{sig}}_{j})}{\sum_{i}N^{\rm{tag}}_{i}},
\label{equ:bsigs2}
\end{equation}
\end{linenomath*}
where $j = 1,2\dots$ is the index for the intervals of $\lmd$ weighting kinematics, and $N^{\rm{sig}}_{-,j}$ is the sum of DT signal yields in the two tag modes within the $j$-th interval.

%%%%%%     Charged     track     %%%%%%%%%%
To select the candidate events, the charged tracks detected in the
main drift chamber (MDC) are required to satisfy $|\cos\theta| <
0.93$, where $\theta$ is the polar angle with respect to the direction
of the $e^{+}$ beam. The distance of closest approach of the charged
tracks to the run-averaged interaction point (IP) must be less than
10\,cm along the beam axis and less than 1\,cm in the perpendicular
plane, except for those tracks used to reconstruct $\kzs$ and
$\lmd$. Particle identification (PID) is achieved by combining the
measurement of specific ionization (dE/dx) and time-of-flight information to compute likelihoods for different particle hypotheses. Protons are distinguished from pions and kaons with the likelihood requirements $\mathcal{L}(p)>\mathcal{L}(K)$ and $\mathcal{L}(p)>\mathcal{L}(\pi)$, while kaons and pions are discriminated from each other by requiring $\mathcal{L}(K)>\mathcal{L}(\pi)$ or $\mathcal{L}(\pi)>\mathcal{L}(K)$, respectively. To improve efficiency, no PID requirements are imposed on the charged pion candidates from the decays of $\lmd$ or $\kzs$.

%%%%%%    $K_{s}^{0}$ $\Lambda$$\Sigma , \omega$  Reconstruction   %%%%%%%%%%
The $\kzs$ and $\lmd$ candidates are reconstructed through their
dominant decays $\kzs \to \pip \pim$ and $\lmd \to p \pim$. The
distances of closest approach of the two candidate charged tracks to
the IP must be within $\pm$20\,cm along the beam direction, with no
requirements imposed in the perpendicular plane. The two charged tracks are
constrained to originate from a common vertex by performing a
vertex fit on the two tracks and requiring the
$\chi^2$ of the fit to be less than 100. A secondary vertex fit is
performed on the daughter tracks of the
surviving $\kzs$ and $\lmd$ candidates,
imposing the additional constraint
that the momentum of the candidate points back to the IP.  The decay
vertex from this secondary vertex fit is required to be on the correct side of the IP and separated
from the IP by a distance of at least twice its fitted resolution. The events with only one pair of charged tracks
satisfying the above requirements are kept, and the fitted momenta of
the $\pip\pim$ and $p\pim$ combinations are used in the further
analysis.
To select $\kzs$ and $\lmd$ candidates, the invariant masses of
$\pip\pim$ and $p\pim$ are required to be in the range
$487<M_{\pi^+\pi^{-}}<511\,\mevcc$ and
$1111<M_{p\pi^{-}}<1121\,\mevcc$, respectively.

\begin{figure}[tp!]
\includegraphics[width=0.45\linewidth]{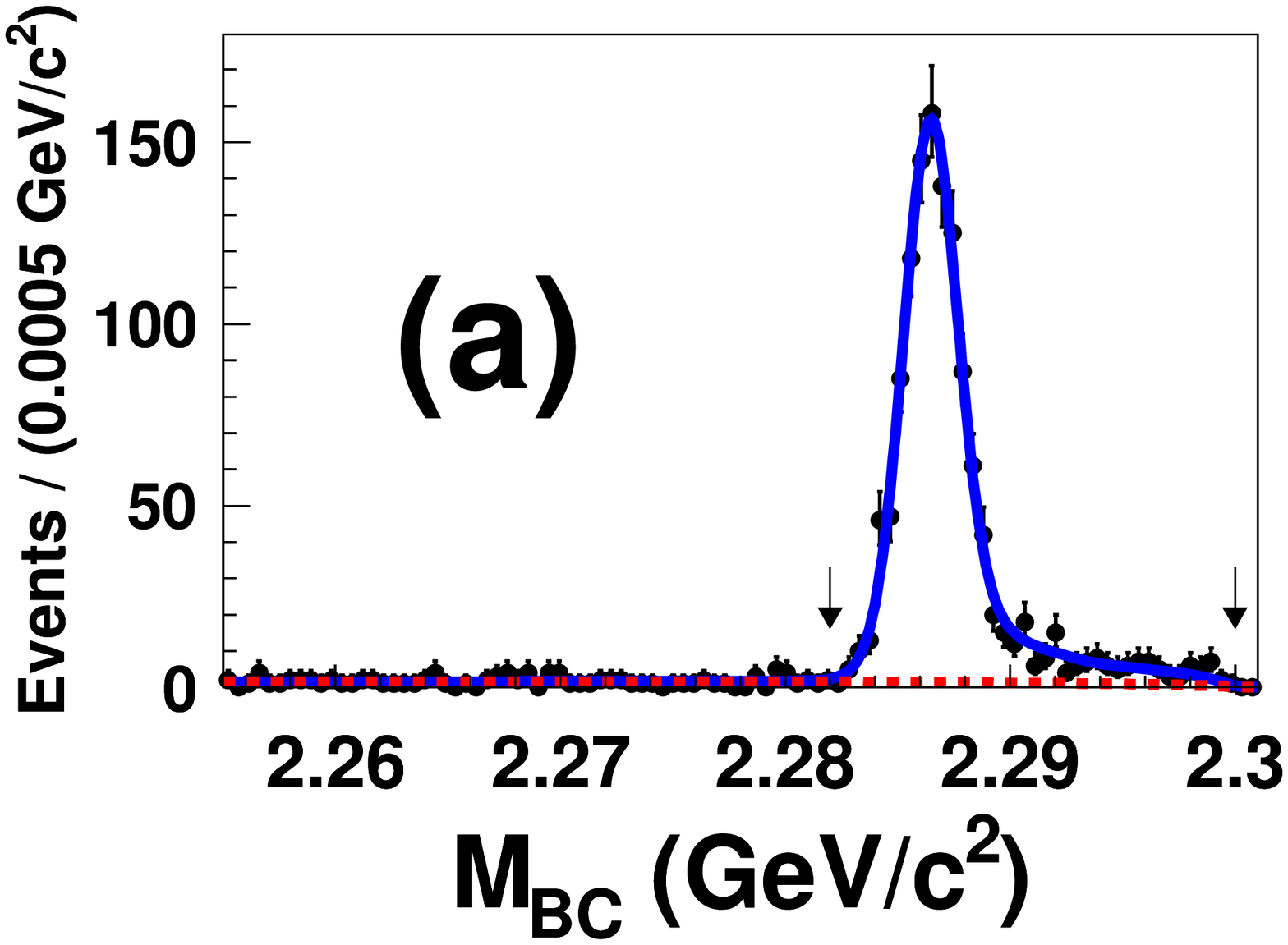}
\includegraphics[width=0.45\linewidth]{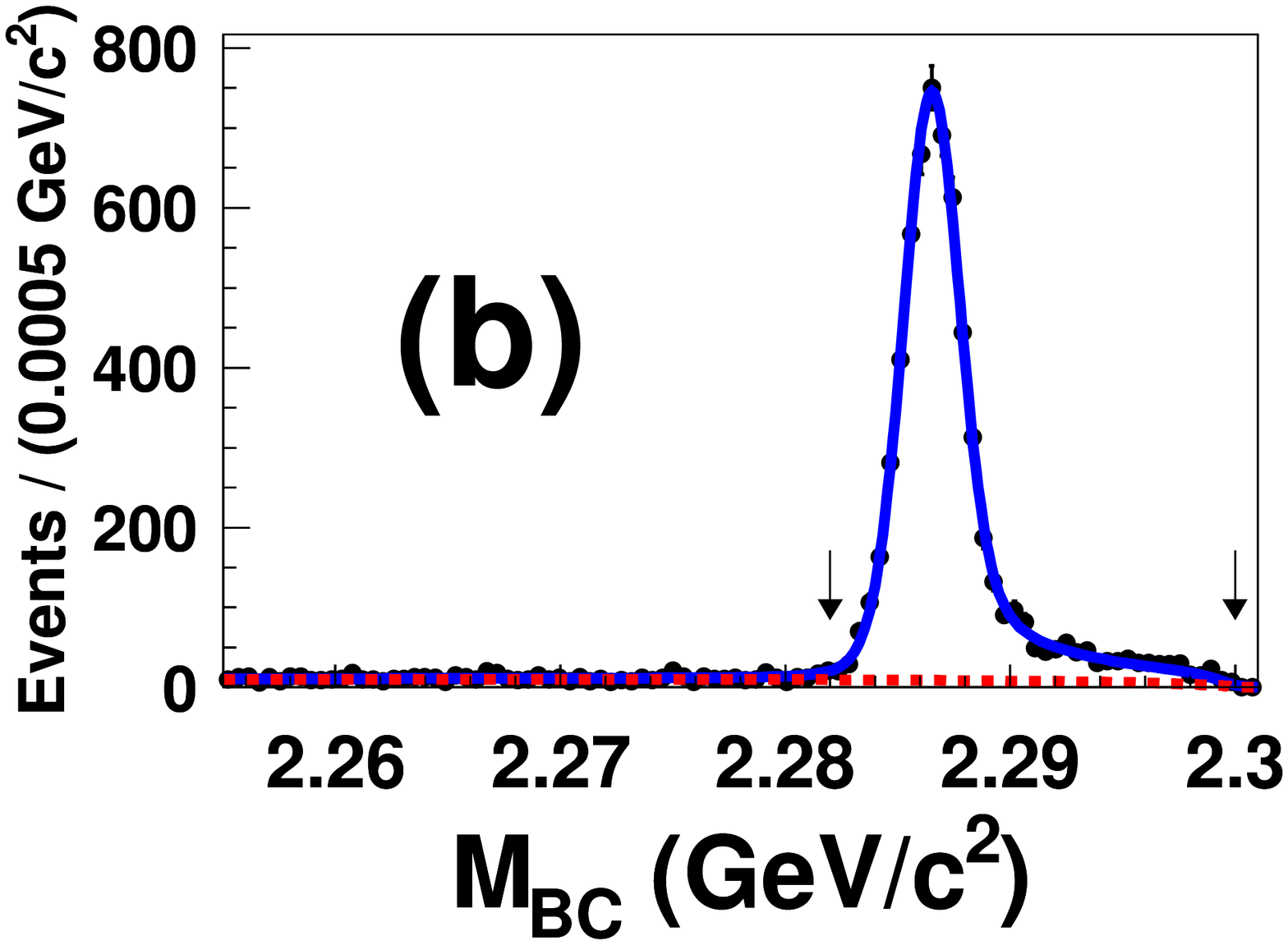}
\caption{Fits to the $\mbc$ distributions of the candidate events for (a)\,$\lcmtopks$ and (b)\,$\lcmtopkpi$ in data. The thick dots stand for the data. The solid curves denote the total fits, while the dotted lines represent the background. The arrows show the signal regions. The description of the fits is given in the text.}
\label{fig:tagmBC}
\end{figure}

To distinguish the tagged $\lcm$ candidates from background, we define
two variables in the $e^+e^-$ rest frame that reflect the conservation of energy and momentum.
The first is the energy difference, $\de \equiv E_{\lcm} - E_{\rm beam}$, where $E_{\lcm}$ is the measured energy of the tagged $\lcm$ candidate and $E_{\rm beam}$ is the beam energy. To suppress combinatorial backgrounds, the mode-dependent $\de$ requirements listed in Table~\ref{tab:tagInfo}, corresponding to $\pm 2.5$ times the resolutions of the fitted $\de$ peaks, are imposed on the tagged $\lcm$ candidates.
The second is the beam-constrained mass of the tagged $\lcm$ candidate, $\mbc \equiv \sqrt{E_{\rm{beam}}^2-|\vec{p}_{\lcm}|^{\ 2} \cdot c^{2}}/c^{2}$, where $\vec{p}_{\lcm}$ represents the momentum of the $\lcm$ candidate.
Figure~\ref{fig:tagmBC} shows the $M_{\rm BC}$ distributions of the
two tag modes, showing clear $\lcm$ signals at the expected
mass. Studies based on MC show that the peaking backgrounds in the tag
modes are negligible. Maximum likelihood fits are performed on these
$\mbc$ distributions to obtain the yields of tagged $\lcm$. The
backgrounds are parameterized by an ARGUS function~\cite{argus} with
endpoint fixed to the beam energy. The signals are described by the
MC-simulated shapes convoluted by Gaussian functions with free widths
to account for the difference of resolutions between data and MC
simulations. The yields for the background and signal are free
parameters in the fits. By subtracting the number of events of the
fitted backgrounds from the total event yields, we obtain the yields of the single tagged $\lcm$, as listed in Table~\ref{tab:tagInfo}.

%%%%%%%%%%%%%%%%%%%%%%%%%%%%%%%%%%%%%%%%%%%%%%%%%%%%%%%%%%%
%%%%%%%%%%%%%%%%%%%%%%%%%%%%%%%%%%%%%%%%%%%%%%%%%%%%%%%%%%%

%%%%%%%%%%%%%%%%%%%%%%%%%%%%%%%%%%%%%%%%%%%%%%%%%%%%%%%%%%%
\begin{table}[tp!]
\begin{center}
\caption{Requirements on $\de$, $\mbc$ and resulting yields $N_i^{\rm tag}$ for the tagged $\lcm$ in data. The uncertainty of $N_i^{\rm tag}$ is statistical only.}
\label{tab:tagInfo}
\begin{tabular}{lccc} \hline \hline
Tag mode $i$ & $\de \,$(GeV) & $\mbc\,(\gevcc)$ & $N_i^{\rm tag}$ \\ \hline
$\lcmtopks$ & $[-0.021,0.019]$ & \multirow{2}{*}{$[2.282, 2.300]$} & $1220 \pm 37$ \\
$\lcmtopkpi$ & $[-0.020, 0.015]$ & & $6088 \pm 85$ \\ \hline \hline
\end{tabular}
\end{center}
\end{table}
%%%%%%%%%%%%%%%%%%%%%%%%%%%%%%%%%%%%%%%%%%%%%%%%%%%%%%%%%%%

\begin{figure}[tp!]
\centering
\includegraphics[width=0.45\textwidth]{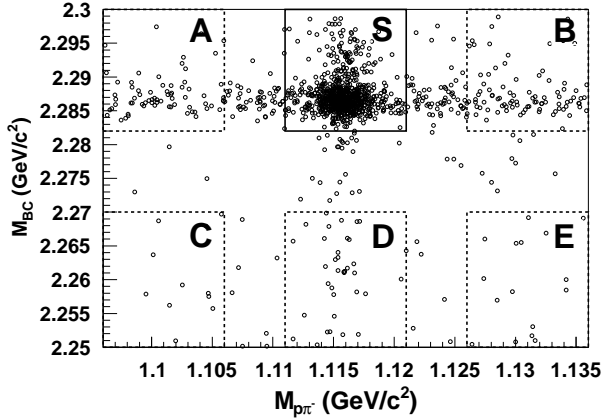}
\caption{Scatter plot of $\mbc$ versus $M_{p\pim}$ of the DT candidates in data. The box labeled S stands for the signal region, while boxes A, B, C, D, and E denote the sideband regions.}
\label{fig:mBCmLmd}
\end{figure}

Then we search for a $\lmd$ candidate among the remaining tracks on the recoiling side of the tagged $\lcm$. The signal yield is determined from the distribution of $\mbc$ versus the invariant mass of $p\pim$ system $M_{p\pim}$ by
\begin{linenomath*}
\begin{equation}
N^{\rm{sig}} = N^{\rm{S}} - \frac{N^{\rm{A}} + N^{\rm{B}}}{2} - f \cdot ({N^{\rm{D}}} - \frac{N^{\rm{C}} + N^{\rm{E}}}{2}),
\label{equ:nLmd}
\end{equation}
\end{linenomath*}
where $N^{\rm{S}}$, $N^{\rm{A}}$, $N^{\rm{B}}$, $N^{\rm{C}}$, $N^{\rm{D}}$ and $N^{\rm{E}}$ represent the numbers of events observed in the regions of S, A, B, C, D and E, as shown in Fig.~\ref{fig:mBCmLmd}. Here the backgrounds due to mis-reconstruction of $\lmd$ are assumed to be flat in the $\mppi$ distribution, which can be estimated from the events in regions A and B. While the peaking backgrounds in the $\mppi$ distribution, which are from non-$\lcp$ decays with $\lmd$ correctly reconstructed, can be estimated using the sideband region of $\mbc$, namely the regions C, D and E. $f$ is the ratio of background area of the signal region over that of the sideband region in the $\mbc$ distribution, which is evaluated to be $0.58 \pm 0.06$ from the fit to the combined $\mbc$ distribution of data for the two tagging modes.
%%%%%%%%%%%%%%%%%%%%%%% // Fig 2 // %%%%%%%%%%%%%%%%%%%%%%%%
%%%%%%%%%%%%%%%%%%%%%%%%%%%%%%%%%%%%%%%%%%%%%%%%%%%%%%%%%%%%
We divide the data into $5 \times 4$  two-dimensional $\pcth$ intervals of $\lmd$ and obtain the net signal yield in each kinematic interval following Eq.~\eqref{equ:nLmd}, as listed in Table~\ref{tab:yieldeff}.
\begin{table}[tp!]
\begin{center}
\scriptsize
\caption{Signal yield and detection efficiency of the inclusive $\lmd$ in each $\pcth$ interval. The uncertainties here are statistical only.}
\label{tab:yieldeff}
\begin{tabular}{c|cccc} \hline \hline
& \multicolumn{4}{c}{$N^{\rm sig}_{-,j}$} \\ \cline{2-5}
$p\,(\gevc)$ & \multicolumn{4}{c}{$|\cth|$} \\ \cline{2-5}
& $[\,0.00,0.20)$ & $[\,0.20,0.40)$ & $[\,0.40,0.65)$ & $[\,0.65,1.00)$ \\ \hline
$[\,0.0,0.3)$&$5.3\scriptscriptstyle{^{+5.1}_{-3.8}}$&$11.4\scriptscriptstyle{^{+5.5}_{-4.2}}$&$9.1\scriptscriptstyle{^{+5.5}_{-4.2}}$&$6.3\scriptscriptstyle{^{+5.4}_{-4.0}}$ \\
$[\,0.3,0.5)$&$59.8\scriptscriptstyle{^{+9.9}_{-8.6}}$&$41.6\scriptscriptstyle{^{+8.9}_{-7.7}}$&$71.9\scriptscriptstyle{^{+10.7}_{-9.5}}$&$33.1\scriptscriptstyle{^{+8.7}_{-7.4}}$ \\
$[\,0.5,0.7)$&$86.7\scriptscriptstyle{^{+10.9}_{-9.7}}$&$72.5\scriptscriptstyle{^{+10.0}_{-8.8}}$&$74.8\scriptscriptstyle{^{+10.1}_{-9.0}}$&$53.9\scriptscriptstyle{^{+9.1}_{-7.9}}$ \\
$[\,0.7,0.9)$&$40.4\scriptscriptstyle{^{+7.8}_{-6.6}}$&$28.3\scriptscriptstyle{^{+6.8}_{-5.6}}$&$44.0\scriptscriptstyle{^{+8.1}_{-6.9}}$&$38.4\scriptscriptstyle{^{+7.9}_{-6.7}}$ \\
$[\,0.9,1.1)$&$6.9\scriptscriptstyle{^{+4.3}_{-3.0}}$&$12.4\scriptscriptstyle{^{+5.0}_{-3.7}}$&$8.3\scriptscriptstyle{^{+4.2}_{-2.9}}$&$5.5\scriptscriptstyle{^{+3.9}_{-2.6}}$ \\ \hline
& \multicolumn{4}{c}{$\varepsilon^{\rm sig}_{j} \, (\%)$} \\ \cline{2-5}
$p\,(\gevc)$ & \multicolumn{4}{c}{$|\cth|$} \\ \cline{2-5}
& $[\,0.00,0.20)$ & $[\,0.20,0.40)$ & $[\,0.40,0.65)$ & $[\,0.65,1.00)$ \\ \hline
$[\,0.0,0.3)$&$8.28 \pm 0.38$&$8.22 \pm 0.37$&$8.01 \pm 0.31$&$4.45 \pm 0.21$ \\
$[\,0.3,0.5)$&$29.03 \pm 0.37$&$28.28 \pm 0.37$&$26.56 \pm 0.33$&$14.98 \pm 0.21$ \\
$[\,0.5,0.7)$&$35.43 \pm 0.32$&$35.00 \pm 0.33$&$33.25 \pm 0.32$&$20.15 \pm 0.25$ \\
$[\,0.7,0.9)$&$39.68 \pm 0.47$&$39.27 \pm 0.50$&$36.56 \pm 0.50$&$23.80 \pm 0.51$ \\
$[\,0.9,1.1)$&$40.82 \pm 0.14$&$40.21 \pm 0.14$&$37.76 \pm 0.12$&$29.97 \pm 0.11$ \\ \hline \hline
\end{tabular}
\end{center}
\end{table}

%%%%%%%%%%%%%%%%%%%%% // Efficiency // %%%%%%%%%%%%%%%%%%%%%%
The efficiencies for detecting a $\lmd$ candidate are estimated from the control samples $\jpsitoll$ and $\jpsitopbkl$, which are selected from a $J/\psi$ on-peak data sample consisting of $(1310.6\pm7.0)\times10^{6}$ $J/\psi$ decays~\cite{jpsi}. In each kinematic interval, the data-driven efficiency is calculated based on a ``tag-and-probe" technique. For $\jpsitoll$, a $\overline{\Lambda}$ is tagged in an event, while for $\jpsitopbkl$, two charged tracks identified as a proton and a kaon are selected. The missing $\Lambda$ is identified by limiting the missing mass within $[1.067,1.155]\,\gevcc$ for $\jpsitoll$ and $[1.093,1.139]\,\gevcc$ for $\jpsitopbkl$. In the tagged event, we search for a $\lmd$ among the remaining tracks and take the detection rate as the efficiency. We partition the control samples into $\pcth$ intervals, and then determine the efficiency in each interval, as listed in Table~\ref{tab:yieldeff}. For these efficiencies, the BF of the intermediate process $\lmd \to p \pi^{-}$ has been included, and the uncertainties are statistical only.
Inserting the numbers of $N^{\rm{tag}}_{i}$ from Table~\ref{tab:tagInfo}, and the numbers of $N^{\rm sig}_{-,j}$ and $\varepsilon^{\rm sig}_{j}$ from Table~\ref{tab:yieldeff} into Eq.~(\ref{equ:bsigs2}), we determine the BF of $\lcptolx$ to be $\mathcal{B}(\lcptolx)=(38.2^{+2.8}_{-2.3})\%$.
The reliability of the analysis method used in this work has been validated by analyzing the inclusive MC sample.

The \CP asymmetry of the decay $\lcptolx$  is obtained by comparing
the separate BFs of the charge conjugate decays, which are
$\mathcal{B}(\lcptolx) = (39.4^{+4.7}_{-3.4})\%$ and
$\mathcal{B}(\lcmtolx) = (37.8^{+3.8}_{-2.9})\%$. The yields and
efficiencies of $\lcptolx$ and $\lcmtolx$ can be found in the
supplemental material~\cite{attach}. The \CP asymmetry is determined
to be $\mathcal{A}_{\CP}=(2.1^{+7.0}_{-6.6})\%$, where the
uncertainty is statistical only.

%%%%%%%%%%%%%% // Systematic Uncertainties // %%%%%%%%%%%%%%%
In the BF measurement with the DT method, systematic uncertainties from the tag side mostly cancel. Other non-canceling systematic uncertainties, which are estimated relative to the measured BF, are discussed below. The limited statistics of the $\lmd$ control samples bring uncertainty to the $\lmd$ efficiency, which is estimated by a weighted root-mean-square\,(RMS) of the statistical uncertainties for different $\pcth$ intervals given in Table\,\ref{tab:yieldeff}. In this analysis, the efficiency for reconstructing a $\lcm$ using the tag modes or finding a $\lmd$ in the $\lcp$ side have been assumed to be independent of the multiplicities of the $\lcp$\,/\,$\lcm$ sides. To evaluate the potential bias of this assumption, we use MC simulation to study the $\lmd$ efficiencies with 2 different tag modes, or the tag efficiencies with and without inclusion of non-$\Lambda$-involved $\lcp$ decays in the signal side. We find the resultant changes on the $\lmd$ efficiency or tag efficiency are at the percent level, which are taken as the systematic uncertainties. The choice of kinematic intervals is varied and the resultant changes on the output BF are examined. The maximum change is quoted as the systematic uncertainty.
The uncertainty due to the fitting procedure of tag yields is studied by altering the signal shape, fitting range and endpoint of the ARGUS function. Potential bias of the background-subtraction procedure in Eq.~\eqref{equ:nLmd} is studied by changing the boundaries of sideband regions and taking the largest difference in the resultant BF as the systematic uncertainty. All of the above systematic uncertainties are summarized in Table~\ref{tab:syserr} and the total uncertainty is determined to be 2.3\% as the sum in quadrature. For the charge asymmetry $\mathcal{A}_{\CP}$, we assume that the systematic uncertainties for the channels of $\lmd$ and $\lmb$ are the same and completely uncorrelated.

%%%%%%%%%%%%%%%%%%%%%%%%%%%%%%%%%%%%%%%%%%%%%%%%%%%%%%%%%%%%
\begin{table}[!htb]
\begin{center}
\footnotesize
\caption{Summary of the relative systematic uncertainties for the BF of $\lcptolx$.}
\label{tab:syserr}
\begin{tabular}{cc} \hline \hline
Source  & Relative uncertainty~(\%) \\\hline
Statistics of the control sample & $0.6$ \\
$\lmd$ efficiency bias & $1.1$ \\
Tag efficiencies bias & $1.6$ \\
Choices of the intervals & $0.5$ \\
Tag yields & $0.9$ \\
Background subtraction & $0.3$ \\ \hline
Total & $2.3$ \\ \hline \hline
\end{tabular}
\end{center}
\end{table}
%%%%%%%%%%%%%%%%%%%%%%%%%%%%%%%%%%%%%%%%%%%%%%%%%%%%%%%%%%%%

%%%%%%%%%%%%%%%%%%%%%% // Summary // %%%%%%%%%%%%%%%%%%%%%%%%
In summary, by analyzing a data sample taken at $\sqrt{s}=4.6\,\gev$
with the BESIII detector, we report the absolute BF of the inclusive
decay of $\lcptolx$ to be
$\mathcal{B}(\lcptolx)=(38.2^{+2.8}_{-2.2}\pm0.8)\%$. The precision of
the BF is improved by a factor of 4 compared to previous
measurements~\cite{pdg1}. This inclusive rate is larger than the
exclusive rate of $(24.5 \pm 2.1)\%$ in PDG~\cite{pdg1}, which
indicates that more than one third of the $\Lambda_c^+$ decays to
$\Lambda$ remain unobserved in experiment. Furthermore, we search for
direct \CP violation in this decay for the first time. The \CP asymmetry is measured to be $\mathcal{A}_{\CP} = (2.1^{+7.0}_{-6.6}\pm1.4)\%$. The precision is limited by statistical uncertainty and no evidence for \CP violation is found.

%%%%%%%%%%%%%%%%%% // Acknowledgements // %%%%%%%%%%%%%%%%%%%
The authors would like to thank Hai-Yang Cheng and Fu-Sheng Yu for useful discussions. The BESIII collaboration thanks the staff of BEPCII and the IHEP computing center for their strong support. This work is supported in part by National Key Basic Research Program of China under Contract No. 2015CB856700; National Natural Science Foundation of China (NSFC) under Contracts Nos. 11335008, 11425524, 11625523, 11635010, 11735014; the Chinese Academy of Sciences (CAS) Large-Scale Scientific Facility Program; the CAS Center for Excellence in Particle Physics (CCEPP); Joint Large-Scale Scientific Facility Funds of the NSFC and CAS under Contracts Nos. U1532257, U1532258, U1732263; CAS Key Research Program of Frontier Sciences under Contracts Nos. QYZDJ-SSW-SLH003, QYZDJ-SSW-SLH040; 100 Talents Program of CAS; INPAC and Shanghai Key Laboratory for Particle Physics and Cosmology; German Research Foundation DFG under Contracts Nos. Collaborative Research Center CRC 1044, FOR 2359; Istituto Nazionale di Fisica Nucleare, Italy; Koninklijke Nederlandse Akademie van Wetenschappen (KNAW) under Contract No. 530-4CDP03; Ministry of Development of Turkey under Contract No. DPT2006K-120470; National Science and Technology fund; The Swedish Research Council; U. S. Department of Energy under Contracts Nos. DE-FG02-05ER41374, DE-SC-0010118, DE-SC-0010504, DE-SC-0012069; University of Groningen (RuG) and the Helmholtzzentrum fuer Schwerionenforschung GmbH (GSI), Darmstadt.
%%%%%    bibliographies       Part                %%%%%%%%%%%%%
%%%%%%%%%%%%%%%%%%%%%%%%%%%%%%%%%%%%%%%%%%%%%%%%%%%%%%%%%%%%%%%%

\clearpage
\onecolumngrid
\appendix

\section*{Supplemental material}

\begin{table*}[h!]
\renewcommand{\arraystretch}{1.5}
\begin{center}
\scriptsize
\caption{Signal yields and detection efficiencies of the inclusive $\lmd$ and $\lmb$ in each $\pcth$ interval. The errors here reflect only statistical uncertainties.}
\label{tab:lmdyieldeff}
\begin{tabular}{c|cccc|cccc} \hline \hline
& \multicolumn{4}{c|}{$N^{\rm sig, \lmd}_{-,j}$} & \multicolumn{4}{c}{$N^{\rm sig, \lmb}_{-,j}$} \\ \cline{2-9}
$p\,(\gevc)$ & \multicolumn{4}{c|}{$|\cth|$} & \multicolumn{4}{c}{$|\cth|$} \\ \cline{2-9}
& $[\,0.00,0.20)$ & $[\,0.20,0.40)$ & $[\,0.40,0.65)$ & $[\,0.65,1.00)$ & $[\,0.00,0.20)$ & $[\,0.20,0.40)$ & $[\,0.40,0.65)$ & $[\,0.65,1.00)$ \\ \hline
$[\,0.0,0.3)$&$4.5^{+3.7}_{-2.4}$&$4.2^{+4.3}_{-3.0}$&$5.9^{+4.4}_{-3.1}$&$6.8^{+4.6}_{-3.3}$&$0.8^{+4.4}_{-3.0}$&$7.2^{+4.3}_{-3.0}$&$3.2^{+4.3}_{-2.9}$&$-0.5^{+3.8}_{-2.4}$ \\
$[\,0.3,0.5)$&$26.8^{+7.2}_{-5.9}$&$17.8^{+6.2}_{-5.0}$&$44.7^{+8.3}_{-7.1}$&$11.5^{+6.0}_{-4.7}$&$33.0^{+7.6}_{-6.4}$&$23.8^{+7.2}_{-5.9}$&$27.2^{+7.5}_{-6.3}$&$21.6^{+7.1}_{-5.8}$ \\
$[\,0.5,0.7)$&$46.8^{+8.3}_{-7.2}$&$35.0^{+7.1}_{-5.9}$&$30.8^{+7.1}_{-5.9}$&$30.8^{+7.1}_{-5.9}$&$39.8^{+7.8}_{-6.6}$&$37.5^{+7.8}_{-6.6}$&$44.0^{+8.0}_{-6.8}$&$23.1^{+6.6}_{-5.3}$ \\
$[\,0.7,0.9)$&$21.5^{+6.0}_{-4.8}$&$14.4^{+5.2}_{-4.0}$&$21.0^{+6.0}_{-4.8}$&$17.9^{+5.9}_{-4.7}$&$18.9^{+5.9}_{-4.6}$&$13.9^{+5.3}_{-4.1}$&$23.0^{+6.2}_{-5.0}$&$20.5^{+6.0}_{-4.8}$ \\
$[\,0.9,1.1)$&$3.0^{+3.2}_{-1.8}$&$7.0^{+3.9}_{-2.7}$&$4.0^{+3.4}_{-2.1}$&$4.5^{+3.7}_{-2.4}$&$3.9^{+3.8}_{-2.5}$&$5.4^{+4.0}_{-2.6}$&$4.3^{+3.4}_{-2.2}$&$1.0^{+2.6}_{-1.2}$ \\ \hline
& \multicolumn{4}{c|}{$\varepsilon_j^{\rm sig, \lmd}(\%)$} & \multicolumn{4}{c}{$\varepsilon_j^{\rm sig, \lmb}(\%)$} \\ \cline{2-9}
$p\,(\gevc)$ & \multicolumn{4}{c|}{$|\cth|$} & \multicolumn{4}{c}{$|\cth|$} \\ \cline{2-9}
& $[\,0.00,0.20)$ & $[\,0.20,0.40)$ & $[\,0.40,0.65)$ & $[\,0.65,1.00)$ & $[\,0.00,0.20)$ & $[\,0.20,0.40)$ & $[\,0.40,0.65)$ & $[\,0.65,1.00)$ \\ \hline
$[\,0.0,0.3)$&$7.98 \pm 0.53$&$8.25 \pm 0.52$&$7.75 \pm 0.44$&$3.85 \pm 0.27$&$8.54 \pm 0.53$&$8.18 \pm 0.52$&$8.24 \pm 0.45$&$4.99 \pm 0.30$ \\
$[\,0.3,0.5)$&$30.38 \pm 0.55$&$29.38 \pm 0.55$&$27.01 \pm 0.48$&$15.53 \pm 0.32$&$27.83 \pm 0.50$&$27.30 \pm 0.50$&$26.16 \pm 0.45$&$14.49 \pm 0.29$ \\
$[\,0.5,0.7)$&$35.92 \pm 0.47$&$35.91 \pm 0.49$&$34.14 \pm 0.47$&$20.49 \pm 0.36$&$35.02 \pm 0.43$&$34.22 \pm 0.44$&$32.49 \pm 0.43$&$19.85 \pm 0.34$ \\
$[\,0.7,0.9)$&$40.15 \pm 0.70$&$39.30 \pm 0.72$&$36.51 \pm 0.71$&$24.05 \pm 0.75$&$39.30 \pm 0.64$&$39.26 \pm 0.68$&$36.60 \pm 0.69$&$23.56 \pm 0.71$ \\
$[\,0.9,1.1)$&$41.34 \pm 0.20$&$40.70 \pm 0.19$&$38.38 \pm 0.17$&$30.72 \pm 0.16$&$40.30 \pm 0.19$&$39.74 \pm 0.16$&$37.17 \pm 0.16$&$29.25 \pm 0.15$ \\ \hline \hline
\end{tabular}
\end{center}
\end{table*}


\begin{thebibliography}{99}

%\bibitem{hqet} Matthias Neubert, Theory Division, CERN, CH-1211 Geneva 23, Switzerland (1996).

\bibitem{Korner:1978ec}
  J.~G.~K\"orner, G.~Kramer and J.~Willrodt,
  %``Weak Decays of the Charmed Baryon C$_0^+$ and the Inclusive Yield of $\Lambda$ and $p$,''
  Phys.\ Lett.\ B {\bf 78}, 492 (1978)
  Erratum: [Phys.\ Lett.\ B {\bf 81}, 419 (1979)].

\bibitem{Asner:2008nq}
  D.~M.~Asner {\it et al.},
  %``Physics at BES-III,''
  Int.\ J.\ Mod.\ Phys.\ A {\bf 24}, S1 (2009).

\bibitem{Cheng:2015iom}
  H.~Y.~Cheng,
  %``Charmed baryons circa 2015,''
  Front.\ Phys.\ (Beijing) {\bf 10}, 101406 (2015).

\bibitem{Cheng:1997xba}
  H.~Y.~Cheng,
  %``A Phenomenological analysis of heavy hadron lifetimes,''
  Phys.\ Rev.\ D {\bf 56}, 2783 (1997).


%\cite{Olive:2016xmw}
\bibitem{pdg1}
  C.~Patrignani {\it et al.} (Particle Data Group),
  %``Review of Particle Physics,''
  Chin.\ Phys.\ C {\bf 40}, 100001 (2016) and 2017 update.


\bibitem{Cheng:1991sn}
  H.~Y.~Cheng and B.~Tseng,
  %``Nonleptonic weak decays of charmed baryons,''
  Phys.\ Rev.\ D {\bf 46}, 1042 (1992)
  Erratum: [Phys.\ Rev.\ D {\bf 55}, 1697 (1997)].

\bibitem{Cheng:1993gf}
  H.~Y.~Cheng and B.~Tseng,
  %``Cabibbo allowed nonleptonic weak decays of charmed baryons,''
  Phys.\ Rev.\ D {\bf 48}, 4188 (1993).

\bibitem{Geng:2017mxn}
  C.~Q.~Geng, Y.~K.~Hsiao, C.~W.~Liu and T.~H.~Tsai,
  %``Charmed Baryon Weak Decays with SU(3) Flavor Symmetry,''
  arXiv:1709.00808 [hep-ph].

\bibitem{Yu:2017zst}
  F.~S.~Yu, H.~Y.~Jiang, R.~H.~Li, C.~D.~L\"{u}, W.~Wang and Z.~X.~Zhao,
  %``Discovery Potentials of Doubly Charmed Baryons,''
  arXiv:1703.09086 [hep-ph].

  \bibitem{Aaij:2017ueg}
  R.~Aaij {\it et al.} (LHCb Collaboration),
  %``Observation of the doubly charmed baryon $\Xi_{cc}^{++}$,''
  Phys.\ Rev.\ Lett.\  {\bf 119}, 112001 (2017).

\bibitem{exp1}
  K.~Abe {\it et al.} (SLAC Hybrid Facility Photon Collaboration),
  %``Lifetimes, Cross-sections and Production Mechanisms of Charmed Particles Produced by 20-{GeV} Photons,''
  Phys.\ Rev.\ D {\bf 33}, 1 (1986).
\bibitem{exp2}
  M.~Adamovich {\it et al.} (Photon Emulsion Collaboration),
  %``Measurement of Charmed Particle Lifetimes and Decay Branching Ratios,''
  Europhys.\ Lett.\  {\bf 4}, 887 (1987).
\bibitem{exp3}
  G.~D.~Crawford {\it et al.} (CLEO Collaboration),
  %``Measurement of baryon production in B meson decay,''
  Phys.\ Rev.\ D {\bf 45}, 752 (1992).


%\bibitem{theo2} G. Crawford \emph{et al.} (CLEO Collaboration), Phys. Rev. D \textbf{45} 752 (1992).

%\bibitem{bes3} M. Ablikim \emph{et al.} (BESIII Collaboration), Nucl. Instrum. Meth. A \textbf{614}, 345321 (2010).

%\bibitem{ckm1} N. Cabibbo, Phys. Rev. Lett. \textbf{10}, 531 (1963).

\bibitem{ckm2} M. Kobayashi and T. Maskawa, Prog. Theor. Phys. \textbf{49}, 652 (1973).

%\cite{Charles:2004jd}
\bibitem{Charles:2004jd}
  J.~Charles {\it et al.} (CKMfitter Group),
  %``CP violation and the CKM matrix: Assessing the impact of the asymmetric $B$ factories,''
  Eur.\ Phys.\ J.\ C {\bf 41} (2005) no.1,  1.  Updates at \url{ckmfitter.in2p3.fr}.
  %doi:10.1140/epjc/s2005-02169-1
  %[hep-ph/0406184].
  %%CITATION = doi:10.1140/epjc/s2005-02169-1;%%
  %1595 citations counted in INSPIRE as of 06 Mar 2018

%\cite{Bona:2005vz}
\bibitem{Bona:2005vz}
  M.~Bona {\it et al.} (UTfit Collaboration),
  %``The 2004 UTfit collaboration report on the status of the unitarity triangle in the standard model,''
  JHEP {\bf 0507}, 028 (2005).
  Updates at \url{www.utfit.org}
  %doi:10.1088/1126-6708/2005/07/028
  %[hep-ph/0501199].
  %%CITATION = doi:10.1088/1126-6708/2005/07/028;%%
  %346 citations counted in INSPIRE as of 06 Mar 2018

\bibitem{cpcharm1} J. Charles, S. Descotes-Genon, X. W. Kang, H. B. Li and G. R. Lu, Phys. Rev. D \textbf{81}, 054032 (2010).

\bibitem{cpcharm2} X. W. Kang and H. B. Li, Phys. Lett. B \textbf{684}, 137 (2010).

%\cite{Kang:2010td}
\bibitem{cpcharm3}
  X.~W.~Kang, H.~B.~Li, G.~R.~Lu and A.~Datta,
  Int.\ J.\ Mod.\ Phys.\ A {\bf 26}, 2523 (2011).

\bibitem{cpcharm4} M. Bobrowski, A. Lenz, J. Riedl and J. Rohrwild, JHEP \textbf{1003}, 009 (2010).
% [arXiv:hep-ph/1002.4794].

%\bibitem{bepc2} C. Zhang \emph{et al.}, Sci. China G \textbf{53}, 2084 (2010)


%\cite{Ablikim:2015nan}
\bibitem{llpair}
  M.~Ablikim {\it et al.} (BESIII Collaboration),
  %``Precision measurement of the integrated luminosity of the data taken by BESIII at center of mass energies between 3.810 GeV and 4.600 GeV,''
  Chin.\ Phys.\ C {\bf 39}, 093001 (2015).
%  doi:10.1088/1674-1137/39/9/093001
  %[arXiv:1503.03408 [hep-ex]].
  %%CITATION = doi:10.1088/1674-1137/39/9/093001;%%
  %30 citations counted in INSPIRE as of 23 Jan 2017
%\bibitem{llpair} M. Ablikim \emph{et al.} Chinese Phys. C \textbf{39}, 093001 (2015).

\bibitem{Ablikim:2015flg}
  M.~Ablikim {\it et al.} (BESIII Collaboration),
  %``Measurements of absolute hadronic branching fractions of $\Lambda_{c}^{+}$ baryon,''
  Phys.\ Rev.\ Lett.\  {\bf 116}, 052001 (2016).


%\cite{Baltrusaitis:1985iw}
\bibitem{mark3}
  R.~M.~Baltrusaitis {\it et al.} (MARK-III Collaboration),
  %``Direct Measurements of Charmed d Meson Hadronic Branching Fractions,''
  Phys.\ Rev.\ Lett.\  {\bf 56}, 2140 (1986).
%  doi:10.1103/PhysRevLett.56.2140
  %%CITATION = doi:10.1103/PhysRevLett.56.2140;%%
  %170 citations counted in INSPIRE as of 23 Jan 2017
%\bibitem{mark3} R.~M.~Baltrusaitis \emph{et al.} [MARK-III Collaboration), Phys. Rev. L \textbf{56}, 2140 (1986).

%\cite{Ablikim:2009aa}
\bibitem{Ablikim:2009aa}
  M.~Ablikim {\it et al.} (BESIII Collaboration),
  %``Design and Construction of the BESIII Detector,''
  Nucl.\ Instrum.\ Meth.\ A {\bf 614}, 345 (2010).
 % doi:10.1016/j.nima.2009.12.050
  %[arXiv:0911.4960 [physics.ins-det]].
  %%CITATION = doi:10.1016/j.nima.2009.12.050;%%
  %332 citations counted in INSPIRE as of 23 Jan 2017

%\cite{Agostinelli:2002hh}
\bibitem{Agostinelli:2002hh}
  S.~Agostinelli {\it et al.} (GEANT4 Collaboration),
  %``GEANT4: A Simulation toolkit,''
  Nucl.\ Instrum.\ Meth.\ A {\bf 506}, 250 (2003).
%  doi:10.1016/S0168-9002(03)01368-8
  %%CITATION = doi:10.1016/S0168-9002(03)01368-8;%%
  %7356 citations counted in INSPIRE as of 23 Jan 2017

%\cite{Jadach:1999vf}
\bibitem{Jadach:1999vf}
  S.~Jadach, B.~F.~L.~Ward and Z.~Was,
  %``The Precision Monte Carlo event generator K K for two fermion final states in e+ e- collisions,''
  Comput.\ Phys.\ Commun.\  {\bf 130}, 260 (2000);
%  doi:10.1016/S0010-4655(00)00048-5
  %[hep-ph/9912214].
  %%CITATION = doi:10.1016/S0010-4655(00)00048-5;%%
  %559 citations counted in INSPIRE as of 23 Jan 2017
  %\cite{Jadach:2000ir}
%\bibitem{Jadach:2000ir}
  %S.~Jadach, B.~F.~L.~Ward and Z.~Was,
  %``Coherent exclusive exponentiation for precision Monte Carlo calculations,''
  Phys.\ Rev.\ D {\bf 63}, 113009 (2001).
  %doi:10.1103/PhysRevD.63.113009
  %[hep-ph/0006359].
  %%CITATION = doi:10.1103/PhysRevD.63.113009;%%
  %280 citations counted in INSPIRE as of 23 Jan 2017

%\cite{Lange:2001uf}
\bibitem{Lange:2001uf}
  D.~J.~Lange,
  %``The EvtGen particle decay simulation package,''
  Nucl.\ Instrum.\ Meth.\ A {\bf 462}, 152 (2001);
%  doi:10.1016/S0168-9002(01)00089-4
  %%CITATION = doi:10.1016/S0168-9002(01)00089-4;%%
  %1592 citations counted in INSPIRE as of 23 Jan 2017
%\cite{Ping:2008zz}
%\bibitem{Ping:2008zz}
  R.~G.~Ping,
  %``Event generators at BESIII,''
  Chin.\ Phys.\ C {\bf 32}, 599 (2008).
  %doi:10.1088/1674-1137/32/8/001
  %%CITATION = doi:10.1088/1674-1137/32/8/001;%%
  %93 citations counted in INSPIRE as of 23 Jan 2017

 %\cite{Kuraev:1985hb}
\bibitem{Kuraev:1985hb}
  E.~A.~Kuraev and V.~S.~Fadin,
  %``On Radiative Corrections to e+ e- Single Photon Annihilation at High-Energy,''
  Sov.\ J.\ Nucl.\ Phys.\  {\bf 41}, 466 (1985)
  [Yad.\ Fiz.\  {\bf 41}, 733 (1985)].
  %%CITATION = SJNCA,41,466;%%
  %722 citations counted in INSPIRE as of 23 Jan 2017

 %\cite{RichterWas:1992qb}
\bibitem{RichterWas:1992qb}
  E.~Richter-Was,
  %``QED bremsstrahlung in semileptonic B and leptonic tau decays,''
  Phys.\ Lett.\ B {\bf 303}, 163 (1993).
%  doi:10.1016/0370-2693(93)90062-M
  %%CITATION = doi:10.1016/0370-2693(93)90062-M;%%
  %51 citations counted in INSPIRE as of 23 Jan 2017

\bibitem{pdg2}
  K.~A.~Olive {\it et al.} (Particle Data Group),
  %``Review of Particle Physics,''
  Chin.\ Phys.\ C {\bf 40}, 100001 (2014).

\bibitem{argus} H. Albrecht \emph{et al.} (ARGUS Collaboration), Phys. Lett. B \textbf{241}, 278 (1990).

\bibitem{jpsi}  M.~Ablikim {\it et al.} (BESIII Collaboration),
  Chin.\ Phys.\ C {\bf 41}, 013001 (2017).

\bibitem{attach} See Supplemental Materials at [URL will be inserted by the publisher] for a summary of the yields of $\lcptolx$ and $\lcmtolx$, as well as the reconstruction efficiencies of $\lmd$ and $\lmb$.

\end{thebibliography}
\end{document}